\documentclass[twocolumn,showpacs,amsmath,amssymb,nofootinbib]{revtex4-1}

\begin{document}

\title{Scalar Polynomial Curvature Invariant\\
Vanishing on the Event Horizon of Any Black Hole Metric\\
Conformal to a Static Spherical Metric}
\thanks{Alberta-Thy 2-17}
\author{David D. McNutt}
\email{ david.d.mcnutt@uis.no}
\affiliation{Faculty of Science and Technology,\\ 
                         University of Stavanger, 
                         N-4036 Stavanger, Norway         }

\author{Don N. Page}
\email{profdonpage@gmail.com}
\affiliation{Department of Physics, University of Alberta,
Edmonton, Alberta T6G 2E1, Canada}

\date{2017 April 15 }

\begin{abstract}
We construct a scalar polynomial curvature invariant that transforms covariantly under a conformal transformation from any spherically symmetric metric. This invariant has the additional property that it vanishes on the event horizon of any black hole that is conformal to a static spherical metric. 
\end{abstract}

\maketitle
The location of a black hole event horizon is generically difficult to find, because the horizon depends upon the future evolution of the spacetime.  However, for certain subclasses of black hole spacetimes, the event horizon can be located more simply.  For example, if the spacetime is stationary, then knowing the hypersurface metric and extrinsic curvature at one time should be sufficient to determine the entire spacetime and hence where the horizon is.

In this particular case, if one knew the Killing vector field that becomes the null generator on the event horizon, one could just find out where the squared norm of this vector vanishes, and that would include the horizon.  However, if one does not know this Killing vector field, the squared norms of the wedge products of $n$ linearly independent gradients of scalar polynomial curvature invariants (where $n$ is the local cohomogeneity of the spacetime) vanish at stationary horizons \cite{Page:2015aia}, so one can use these invariants to find locations that include the horizon.

For spacetimes that are not stationary, there is not a general procedure of this form.  However, since the location of the event horizon is a conformal invariant, it can be found by the procedure above for any black hole spacetime metric that is conformal to a stationary metric {\it if} the conformal factor is known that transforms the spacetime metric to the corresponding stationary metric.  Nevertheless, if the conformal factor is {\it not} known, one may not have a clear way to locate the horizon, analogous to the problem of locating the event horizon of a stationary metric when one does not know the Killing vector that becomes null on the horizon.

One way to attempt to locate the horizon of a black hole metric that is conformal to a stationary metric is to search for a scalar polynomial curvature invariant that not only vanishes on the horizon of the stationary metric (as do the squared norms of the wedge products of gradients of invariants described above) but that also remains zero on the horizon under a conformal transformation of the metric.  In particular, we would like an invariant that not only vanishes on a stationary horizon but also under a conformal transformation transforms as a power of the conformal factor, with no terms from derivatives of the conformal factor, which could be nonzero on the event horizon.

Unfortunately, the specific invariants listed explicitly in \cite{Page:2015aia} that vanish on stationary horizons pick up derivative terms when one makes a generic conformal transformation of the metric, so they do not remain zero on the horizon of a generic black hole spacetime that is conformal to a stationary metric.  Perhaps there are examples that do work in general, but we have not yet found them.

In this paper, we solve the more restricted problem of finding a scalar polynomial curvature invariant that vanishes on the horizon of any black hole metric that is conformal to a 4-dimensional spherically symmetric static spacetime, with the conformal factor being any smooth function over the spacetime that does not vanish at the event horizon.

For this purpose, we start with a list of scalar polynomial curvature invariants obtained from the Weyl tensor and its first covariant derivative, all but $I_{3a}$ of which are given in  \cite{Page:2015aia}:

\begin{equation}
 I_1 \equiv C_{abcd}\;C^{abcd}
 \label{defi1} \;,
\end{equation}
\begin{equation}
 I_3 \equiv C_{abcd;e}\;C^{abcd;e}
 \label{defi3}\;,
 \end{equation}
\begin{equation}
 I_{3a} \equiv C_{abcd;e}\;C^{ebcd;a}
 \label{defi4}\;,
\end{equation}
\begin{equation}
 I_5 \equiv I_{1;e}I_1^{\ ;e}
 \label{defi5}\;.
\end{equation}

Consider a 4-dimensional spherically symmetric static black hole spacetime that is smooth on the event horizon, has a nonzero surface gravity $\kappa$, has $2\pi r$ as the circumference of the 2-sphere of the spherical symmetry passing through each event, and admits a Killing vector field $\partial/\partial t$ which is timelike outside the event horizon but null on the event horizon at $r = r_H$. For such a black hole spacetime, replacing $t$ by $-i\tau$ gives a Euclidean-signature metric in which $\tau$ becomes an angular variable with period $2\pi/\kappa$, and for each location on the 2-spheres of the spherical symmetry, the horizon is replaced by a regular center at $r = r_H$ in the $(r,\tau)$ plane where the proper circumference of the angular variable $\tau$ goes to zero. In a smooth orthonormal frame at this center, the gradient of any smooth scalar polynomial curvature invariant vanishes, as well as the first covariant derivative of any curvature tensor such as the Weyl tensor.  Therefore, the invariants $I_3$, $I_{3a}$, and $I_5$ vanish on the event horizon of any 4-dimensional spherically symmetric static black hole spacetime that is smooth on the event horizon and has a nonzero surface gravity $\kappa$, and they continue to vanish when one takes the limit of zero surface gravity, so it is unnecessary to make the restriction to $\kappa \neq 0$.

However, under a conformal transformation of the metric, $I_3$, $I_{3a}$, and $I_5$ do not transform just by being multiplied by powers of the conformal factor, but they also pick up first derivatives of the conformal factor, which generically are nonzero on the event horizon and hence make $I_3$, $I_{3a}$, and $I_5$ nonzero at the event horizon of a generic metric conformal to a static spherically symmetric black hole metric.  Nevertheless, we have found a particular combination of the invariants listed above that under a conformal transformation of the metric does transform purely by being multiplied by a negative power of the conformal factor: 
\begin{equation}
 J_4 \equiv 6 I_1 I_3 -16 I_1 I_{3a} +  I_5.
 \label{J}
\end{equation}
The invariant $J_4$ vanishes on the horizon of a static spherically symmetric metric. Since $J_4$ transforms by being multiplied by a negative power of the conformal factor under a conformal transformation, so long as the conformal factor does not vanish on the event horizon, the scalar polynomial curvature invariant $J_4$ is zero on the event horizon for any smooth 4-dimensional metric conformal to a static spherically symmetric black hole spacetime.

In particular, if we write such a metric as
\begin{align}
 d\hat{s}^2 = e^{2U(t,r,\theta,\phi)}
 [&-e^{2\psi(r)}(1-2M(r)/r)dt^2
  \nonumber\\
  &+ (1-2M(r)/r)^{-1}dr^2
  \nonumber\\
  &+ r^2(d\theta^2 + \sin^2{\theta}d\phi^2)]
 \label{metric}\;,
\end{align}
then
\small 
\begin{align}
&\hat{J}_4 = \frac{2^5}{3} \frac{e^{-10U}}{r^{15}}(r-2M) & \nonumber \\
& \times  [2 M \psi_{,r}^2 r^2 - \psi_{,r}^2 r^3 + 2M \psi_{,r,r} r^2 - \psi_{,r,r}r^3 + 3\psi_{,r} M_{,r} r^2 & \nonumber \\
& + M_{,r,r}r^2 - 5M \psi_{,r}r + \psi_{,r}r^2   - 4M_{,r} r + 6M ]^2 & \nonumber \\
&\times [ 4r^3 \psi_{,r}\psi_{,r,r} M - 2r^4 \psi_{,r}\psi_{,r,r} + 2 \psi_{,r}^2 r^3 M_{,r} + 2r^3 \psi_{,r,r,r}M & \nonumber \\
& - r^4 \psi_{,r,r,r} + 5r^3 \psi_{,r,r} M_{,r}  + 3r^3 \psi_{,r} M_{,r,r} + 2M \psi_{r}^2 r^2 - 2\psi_{,r}^2 r^3 & \nonumber  \\
& - 3M \psi_{,r,r}r^2 - \psi_{,r,r}r^2 + r^3 M_{,r,r,r} - 2\psi_{,r} M_{,r} r^2 - 3M_{,r,r}r^2  & \nonumber \\
& + \psi_{,r}r^2 + 6M_{,r} r - 6M]^2, & \label{Jformetric}\;,
\end{align}

\normalsize
\noindent which vanishes at the event horizon at $r = 2M(r)$.

For a generic non-vacuum metric $g_{ab}$ which is not conformal to a static spherically symmetric metric, and a conformally transformed metric $\hat{g}_{ab} = e^{2U}g_{ab}$, the invariant $\hat{J}_4$ for the metric $\hat{g}_{ab}$ is related to the invariant $J_4$ for the metric $g_{ab}$ by
\begin{equation}
 \hat{J}_4 = e^{-10U}(J_4 + A^a U_{;a})
 \label{Jtransform}\;,
\end{equation}
where
\begin{equation}
A_a = -4 I_{1,a} - 16 W^e_{~ae} + 16 W_{a~e}^{~e} + 64 \bar{W}^e_{~ea} 
\label{A}\;,
\end{equation}
\noindent and 
\begin{eqnarray}
& W^a_{~bc} = C_{bdef;c}C^{adef} & \label{Weqn} \\
& \bar{W}^a_{~bc} = C_{bdce;f}C^{fdae}. & \label{Wbeqn}
\end{eqnarray} 

When $g_{ab}$ is a vacuum metric (including a possible cosmological constant) or is spherically symmetric (and hence also when $g_{ab}$ is conformal to a spherically symmetric metric), it can be shown that $A^a = 0$ by direct calculation. One may use the Newman-Penrose (NP) formalism \cite{Stephani:2003tm} to express the Weyl tensor in terms of the NP curvature scalars $\Psi_{i},~i \in [0,4]$, and the covariant derivative of the Weyl tensor in a concise form. Constructing the co-vector terms in $A_a$ and adding them together shows that $A_a$ will vanish for all four-dimensional spacetimes. 

\begin{section}{Examples in 4D}

It is possible to generate new solutions of the Einstein equation by applying a conformal transformation to a known solution \cite{Stephani:2003tm}, and solving the ensuing differential equations for a particular conformal factor. While conformally-flat solutions have been explored (chapter 37 of \cite{Stephani:2003tm}), less is known about generation of new solutions from conformally non-flat solutions. Starting from a Ricci-flat but conformally non-flat solution, conformally related solutions which admit a perfect fluid were initially studied \cite{Newman:1985, vandenBergh,vandenBergh:2011,Hansraj:2012}. Similar approaches have led to physically acceptable perfect fluid solutions \cite{LL2008}, and viscous fluid solutions \cite{CarotTupper}. Solutions generated by non-Ricci-flat solutions exist as well, which are of Type N and III respectively, satisfying the Einstein-Maxwell equations \cite{Hruska}.

We are interested in black hole solutions generated by a conformal transformation from a static spherically symmetric metric. One of the first examples of such a solution is the Sultana-Dyer metric which is conformally related to the Schwarzschild solution:

\begin{eqnarray}
 ds^2 &=& t^4 \biggl[ - \left( 1 - \frac{2m}{r} \right) dt^2 + \frac{4m}{r} dt dr \\
& & +  \left( 1 - \frac{2m}{r} \right) dr^2 + r^2 (d\theta^2 + \sin^2 \theta d\phi^2)\biggl], \nonumber
\end{eqnarray}

\noindent which describes an expanding black hole in the asymptotic background of the Einstein-de Sitter universe \cite{SD2005} with a matter content described by a two-fluid source: null fluid and pure dust. The event horizon for this solution is now a conformal Killing horizon which is the image of the Killing horizon of the Schwarzschild black hole under a conformal transformation.

Although this is hidden by the choice of coordinates, the Sultana-Dyer metric belongs to a subclass of metrics, called the Thakurta metrics \cite{Thakurta}, which arise from a conformal transformation of the Kerr solution with a conformal factor dependent on the Boyer-Lindquist time coordinate. Setting the rotation parameter to zero, the metric is of the form \cite{MMZ2016}

\begin{eqnarray}
& ds^2 = a^2(\eta) \left[ -\left( 1 - \frac{2m}{r}\right) d \eta^2 + \frac{dr^2}{ 1 - \frac{2m}{r}}+r^2 d \Omega^2 \right]. & 
\end{eqnarray} 
\noindent With a cosmological time $t$ such that $dt = a(\eta) d \eta$, the metric becomes 

\begin{eqnarray}
ds^2 =-\left( 1 - \frac{2m}{r}\right) d t^2 + \frac{a^2 dr^2}{ 1 - \frac{2m}{r}}+a^2r^2 d \Omega^2. \label{metric:Thakurta} 
\end{eqnarray} 

This class of metrics intersects with another class of metrics, namely the generalized McVittie  (or gMcVittie) metrics where the mass parameter is now a function of the time coordinate \cite{M1933,MGM2015, GFdSA2012}. Writing \eqref{metric:Thakurta} in terms of areal radius coordinates $R = ar$, the form of the metric is identical to (9) in \cite{GFdSA2012}; it is therefore a gMcVittie spacetime with $m(t) = ma(t)$. 

The analysis of \cite{MGM2015, GFdSA2012} does not carry over to \eqref{metric:Thakurta} due to the assumption $\dot{m}(t)/ m(t) < \dot{a}(t) / a(t)$,  whereas equality holds in \eqref{metric:Thakurta}. The analysis of a special case of this metric \cite{Faraoni2009} with $a(t) \sim t^\frac23$ describes a universe filled with dust and a singularity at $r=2m$. This metric is distinct from the the Sultana-Dyer metric \cite{CD2010}. 

In \cite{MMZ2016} the analysis of the causal structure of \eqref{metric:Thakurta} is completed and particular examples of scale factors are studied. It is shown that unbounded scale factors yield solutions describing inhomogeneous expanding universes with no horizon, a singular surface at $r=2m$, and potentially null singularities at a finite proper time in the future. For a bounded scale factor with rapidly vanishing time derivatives, the resulting spacetime represents an event horizon for which the analytical extensions yield black hole or white hole regions. The black hole solutions represent dynamical accreting black holes, which at late times decouple from the cosmological expansion and cease to accrete, and hence exist in effectively static bubbles of vacuum.

\end{section}

\begin{section}{Spherically Symmetric Static Spacetimes in Higher Dimensions}

The invariant $J_4$ can be generalized to higher dimensions; however, due to the trace terms $C_{acde} C^{bcde}$ some modification to the coefficients of $I_1, I_3$, and $I_5$ are required to produce an invariant $J_n$ that actually vanishes on the event horizon of any metric that is conformal to the $n$-dimensional static spherically symmetric metric, with a nonzero conformal factor on the horizon.

Using the algebraic Bianchi identities, the transformation rules for the scalar polynomial curvature invariants under a conformal transformation are 
\small 
\begin{eqnarray}
\tilde{I}_1 &=& e^{-4U} I_1, \label{I1transf} \end{eqnarray}
\begin{eqnarray}\tilde{I}_3 &=& e^{-6U} [ I_3 - (2 I_{1,a} +8 W^e_{~ae} - 8 W_{a~e}^{~e} ) U^{,a} + 8 U_{,a}U^{,a} I_1 \nonumber \\
&&+ (4n-8) U^{,a}U_{,e} C_{abcd} C^{ebcd}], \label{I3transf}\end{eqnarray}
\begin{eqnarray}\tilde{I}_{3a} &=& e^{-6U} [ I_{3a} - ( 2 W^e_{~ae} + I_{1,a} - 2 W_{a~e}^{~e} + 4 \bar{W}^e_{~ea} ) U^{,a}   \nonumber \\
&& + 4 U_{,a}U^{,a} I_1 + (n-1)U^{,a}U_{,e} C_{abcd} C^{ebcd}], \label{I3atransf} \end{eqnarray}
\begin{eqnarray} \tilde{I}_5 &=& e^{-10U} [ I_3 - 8 I_{1,a} U^{,a} + 16 U_{,a}U^{,a} I_1], \label{I5transf}
\end{eqnarray}

\noindent where $W^a_{~bc}$ and $\bar{W}^a_{~bc}$ are defined in equations \eqref{Weqn} and \eqref{Wbeqn}. From these we may solve two linearly independent equations for two unknowns to eliminate $U_{,a}U^{,a}$ and $U_{,a} U^{,b}$ from the expression. This yields an invariant that transforms covariantly under a conformal transformation for an $n$-dimensional  manifold: 

\begin{equation}
J_n = 2(n-1) I_1 I_3  - 8(n-2) I_1 I_{3a} +  (n-3)I_5. \label{Jn}
\end{equation}

%

For a generic non-vacuum metric $g_{ab}$, which is not conformal to a static spherically symmetric metric, and for a conformally transformed metric $\hat{g}_{ab} = e^{2U}g_{ab}$, the invariant $\hat{J}_n$ for the metric $\hat{g}_{ab}$ is related to the invariant $J_n$ for the metric $g_{ab}$ by
\begin{equation}
 \hat{J}_n = e^{-10U}(J_n + A^a U_{;a} )
 \label{Jntransform}\;,
\end{equation}
where
\begin{align}
& A_a = -4(n-3) I_{1,a} -  16(W^e_{~ae} -W_{a~e}^{~e}) +  32(n-2)\bar{W}^e_{~ea}, & \label{A}  
\end{align}
\noindent where $W^a_{~bc}$ and $\bar{W}^a_{~bc}$ are defined in equations \eqref{Weqn} and \eqref{Wbeqn}.  If $g_{ab}$  describes a vacuum (possibly with a cosmological constant), or is conformal to a spherically symmetric metric, then $A_a$ vanishes automatically.  In the case of a vacuum metric, this can be proven using the Bianchi identities $$C_{a[bcd]} = 0 \text{ and } C_{ab[cd;e]}=0,$$ which give the following identities:
\begin{eqnarray}
W_{a~e}^{~e}=0,~ 4W^e_{~ae} = I_{1,a},\text{ and } 8 \bar{W}^e_{~ea} = I_{1,a}.
\end{eqnarray}
\noindent Substituting these identities causes the vector in \eqref{A} to vanish. 

While in 4D, the vector $A^a$ vanishes for any spacetime, it is not clear whether $A^a = 0$ for a generic higher dimensional spacetime. This vector appears to vanish for a large class of spacetimes, notably $A^a$ vanishes for the Kerr-NUT-(Anti)-de Sitter solution \cite{generalkerr}, the rotating black ring solution \cite{brwands}, and the supersymmetric black ring solution \cite{SBR}, all of which admit stationary regions containing event horizons. Both of the black ring solutions are of algebraic type ${\bf I}_i$ according to the alignment classification \cite{brwands, CHDG}, and the supersymmetric black ring is non-vacuum. This suggests that the invariant $J_n$ transforms in a covariant manner for a larger class of metrics than the spherically symmetric metrics or vacuum metrics. We expect any proof that $A_a$ generically vanishes in higher dimensions will rely on the identities $C^a_{~bac}=0$ and $C_{a[bcd]}=0$, possibly using a decomposition of the Weyl tensor that generalizes the NP formalism \cite{HDWeyl}. Due to the difference in the cohomogeneity for such stationary solutions, the invariant $J_n$ is no longer assured to detect the event horizon.

To illustrate the use of $J_n$, consider a conformal transformation on the five-dimensional static spherically symmetric metric \cite{Carames:2009} 
\begin{eqnarray} & d\hat{s}^2 = e^{-2 U(t,r,\theta_1,\theta_2,\phi)} [-u(r) dt^2 + v(r) dr^2 \nonumber \\ & + r^2 d \theta_1^2 + r^2 \sin \theta_1 d \theta_2  + r^2 \sin \theta_1 \sin \theta_2 d \phi^2], &  \\
& u(r) = e^{\psi(r)}\left(1-\frac{2M(r)}{r^2}\right), v(r) = \left(1-\frac{2M(r)}{r^2}\right)^{-1}. \nonumber  &
\end{eqnarray}

\noindent Then the following invariant 

\begin{eqnarray} J_5 = 8I_1 I_3 - 24 I_1 I_{3a} +  2 I_5  \end{eqnarray}

\noindent will transform in the desired manner and vanish when $ r^2 = 2M(r)$.  Relative to the coordinate system, $J_5$ takes the form


\begin{eqnarray}
&\hat{J}_5 = 2^5 \cdot \frac{e^{-10U}}{r^{20}}(r^2-2M) & \nonumber \\
& \times  [\psi_{,r}^2 r^4 + \psi_{,r,r}r^4+2\psi_{,r}^2r^2 M -2\psi_{,r,r}r^2 M - 3 \psi_{,r} M_{,r} r^2 & \nonumber \\& - \psi_{,r}r^3  + 8 \psi_{,r} M r - M_{,r,r} r^2 + 6 M_{,r} r - 12 M ]^2 & \nonumber \\
&\times [ -2 \psi_{,r,r} \psi_{,r} r^5 + 4 \psi_{,r,r} \psi_{,r} M r^3 + 2 \psi_{,r}^2 M_{,r} r^3 - 2 \psi_{,r}^2 r^4 & \nonumber \\
& - \psi_{,r,r,r} r^5 + 5 \psi_{,r,r}M_{,r} r^3 - \psi_{,r,r} r^4 + 2 \psi_{,r,r,r} M r^3 & \nonumber \\
& + 3 \psi_{,r} M_{,r,r} r^3 - 8 \psi_{,r,r} M r^3 -8 \psi_{,r} M_{,r} r^2 + \psi_{,r} r^3 & \nonumber \\
& +M_{,r,r,r} r^3 + 8 \psi_{,r} M r - 6M_{,r,r} r^2 + 18 M_{,r} r - 24 M  ]^2, & \label{J5formetric}\;,
\end{eqnarray}
\noindent which vanishes on the event horizon $r^2=2M$.
\end{section}

\begin{section}{Conclusions}
By studying the effect of a conformal transformation on the spherically symmetric metrics, we have found an invariant, $J_n$, for which the effect of a conformal transformation causes the invariant to be multiplied by a power of the conformal factor. For any metric conformally related to a static spherically symmetric metric admitting an event horizon, $J_n$ will detect the event horizon of the original metric \cite{Page:2015aia}, and hence will detect the horizon of the event horizon of the conformally related solution. 

In 4D this invariant transforms in this manner for any spacetime, while in higher dimensions we have argued that this invariant transforms covariantly under a conformal transformation for a larger class of spacetimes than vacuum or spherically symmetric metrics. By direct computation it has beeen verified that $J_n$ transforms covariantly for the 5D Kerr-NUT-(Anti)-de Sitter \cite{generalkerr}, and the stationary regions of the rotating black ring solution and supersymmetric black ring \cite{brwands, SBR, CHDG}. However, due to the difference in cohomogeneity, this invariant is no longer assured to detect the horizon of these stationary solutions. 


Despite its limitations, the invariant provides a helpful tool for determining the event horizon for dynamical black hole solutions that are conformally related to static spherically symmetric black hole solutions, such as the Schwarzschild solution \cite{SD2005,Thakurta,MMZ2016}. Motivated by the Thakurta metric \cite{Thakurta}, it would be worthwhile to construct another functionally independent invariant with the same transformation property as $J_n$, say $K_n$. Using $I_1$, $J_n$, and $K_n$, one may apply the results of \cite{Page:2015aia} to construct an invariant that detects the event horizons of any black hole conformally related to Kerr, or more generally the Kerr NUT-(Anti)-de Sitter solution \cite{generalkerr}.
\end{section}

\begin{section}*{Acknowledgements}  
 
This work was supported through the Research Council of Norway, Toppforsk grant no. 250367: Pseudo-
Riemannian Geometry and Polynomial Curvature Invariants: Classification, Characterisation and Applications (D.M.), and by the Natural Sciences and Engineering Research Council of Canada (D.P.).

\end{section}
%
%

\end{document}